\documentclass{aa}

\usepackage[varg]{txfonts}

\usepackage{color,amsmath}

\bibpunct{(}{)}{;}{a}{}{,} 

\title{The coronal energy input from magnetic braiding}

\author{A. R. Yeates\inst{\ref{inst1}}
\and F. Bianchi\inst{\ref{inst1}}
\and B. T. Welsch\inst{\ref{inst2}}
\and P. J. Bushby\inst{\ref{inst3}}}

\institute{Department of Mathematical Sciences, Durham University, Durham, DH1 3LE, UK \email{anthony.yeates@durham.ac.uk}\label{inst1}
\and
Space Sciences Laboratory, University of California, Berkeley, CA 94720, USA\label{inst2}
\and
School of Mathematics and Statistics, Newcastle University, Newcastle Upon Tyne, NE1 7RU, UK\label{inst3}
}

\date{Received ?? / Accepted ??}

\abstract{We estimate the energy input into the solar corona from photospheric footpoint motions, using observations of a plage region by the Hinode Solar Optical Telescope.  Assuming a perfectly ideal coronal evolution, two alternative lower bounds for the Poynting flux are computed based on field line footpoint trajectories, without requiring horizontal magnetic field data. When applied to the observed velocities, a bound based solely on displacements between the two footpoints of each field line is tighter than a bound based on relative twist between field lines. Depending on the assumed length of coronal magnetic field lines, the higher bound is found to be reasonably tight compared with a Poynting flux estimate using an available vector magnetogram. It is also close to the energy input required to explain conductive and radiative losses in the active region corona. Based on similar analysis of a numerical convection simulation, we suggest that observations with higher spatial resolution are likely to bring the bound based on relative twist closer to the first bound, but not to increase the first bound substantially. Finally, we put an approximate upper bound on the magnetic energy by constructing a hypothetical ``unrelaxed'' magnetic field with the correct field line connectivity.}

\keywords{Sun: corona - Sun: granulation - Sun: magnetic fields - Sun: photosphere}

\begin{document}

\maketitle

\section{Introduction}

One of the central questions in solar physics is how the Sun's magnetic field transmits energy through the photosphere to maintain coronal temperatures in excess of a million kelvin. Currently, two broad classes of mechanism are favored: wave heating and magnetic reconnection \citep[see, for example, the reviews by][]{Klimchuk2006a,Reale2010b,Parnell2012f}. Here we focus on heating by magnetic reconnection, and in particular on the magnetic braiding scenario \citep{Parker1972,Parker1983ar}. Parker proposed that convective motions in the photosphere will shuffle the footpoints of coronal magnetic field lines, causing the field lines to become entangled, or braided. Crucially, this will lead to locally intense magnetic gradients in the corona, allowing energy to be released in many small reconnection events, now known as ``nanoflares'' \citep{Parker1988ax}. The heating of the atmosphere is suggested to result from this continual energy release. Rather than braiding of flux tubes around one another, it is also possible for photospheric motions to inject energy by twisting individual flux tube footpoints \citep{Sturrock1981a, Zirker1993b}, although this may be less efficient than braiding \citep{Berger1991q}.  Recent studies indicate that the particular braiding pattern may have a significant effect on the resultant heating in the corona \citep{Berger2009, WilmotSmith2011}.

Coronal heating mechanisms must account for combined conductive and radiative losses from the active region corona of about $10^7\,{\rm erg}\,{\rm cm}^{-2}\,{\rm s}^{-1}$ \citep{Withbroe1977}. To determine whether the rate of energy input by photospheric  braiding motions is sufficient to supply this, consider the rate of change of magnetic energy
\begin{equation}
W=\int_V\frac{B^2}{8\pi}\,d^3x
\label{eqn:energy}
\end{equation}
in a coronal volume $V$, which is given by
\begin{equation}
\frac{dW}{dt}= -\frac{1}{4\pi}\int_V{\bf E}\cdot\nabla\times{\bf B}\,d^3x - \frac{1}{4\pi}\oint_{\partial V}{\bf E}\times{\bf B}\cdot{\bf n}\,d^2x.
\label{eqn:poynting}
\end{equation}
The first term on the right-hand side of \eqref{eqn:poynting} represents the volume dissipation of magnetic energy in the corona, while the second is the Poynting flux through the boundary of $V$. We are interested in the Poynting flux through the photospheric boundary $S_0$, which may be written
\begin{align}
\frac{1}{4\pi}\int_{S_0}{\bf E}\times{\bf B}\cdot{\bf e}_z\,d^2x &= \frac{1}{4\pi}\int_{S_0}v_z(B_x^2 + B_y^2)\,d^2x \nonumber\\
&\qquad - \frac{1}{4\pi}\int_{S_0}B_z(v_xB_x + v_yB_y)\,d^2x,
\label{eqn:poyntingS0}
\end{align}
where we have assumed an ideal Ohm's law ${\bf E}=-{\bf v}\times{\bf B}$. 
Determining this quantity from observations requires  both vector velocity and vector magnetic field data, which remain challenging to obtain at high cadence and high resolution. Our bounds assume $v_z=0$, so can not be applied to regions with significant flux emergence. The idea  is to estimate the last term in \eqref{eqn:poyntingS0} from just $v_x$, $v_y$ and $B_z$, without needing to know $B_x$ or $B_y$. 

Our bounds assume that the coronal magnetic field evolves ideally during the braiding motions, without dissipation. As a result, if we were to move the photospheric footpoints for longer and longer times, we would accumulate more and more energy in the corona. Previous studies have used numerical MHD simulations \citep[e.g.,][]{Mikic1989,Hendrix1996,Galsgaard1996,Gudiksen2002,Bingert2011} or reduced-MHD simulations \citep{Rappazzo2008} driven by photospheric footpoint motions to determine the level at which the energy input saturates. These models suggest that the braiding process leads to heating rates comparable to that of \citet{Withbroe1977}, although the simulations necessarily have coronal dissipation orders of magnitude too high. Our approach is intended to complement these studies by estimating the energy input by a perfectly ideal evolution.

\citet{Parker1983ar} made the simple estimate that $dW/dt\approx 10^7\,{\rm erg}\,{\rm cm}^{-2}\,{\rm s}^{-1}$ by the following argument. Start with a vertical magnetic field of strength $B=100\,{\rm G}$ between $z=0$ and $z=h=100\,{\rm Mm}$. If we displace the footpoint of a flux tube through distance $vt$, then the tube will gain a transverse flux density $B_h=Bvt/h$. The displacement does work against the magnetic stress $BB_h/(4\pi)$, so the power input will be $dW/dt=vBB_h/(4\pi)=v^2B^2t/(4\pi h)$. Using the speed $v=0.4\,{\rm km}\,{\rm s}^{-1}$ suggested by bright-point observations \citep{Smithson1973}, and an assumed time of $t=1\,{\rm day}$ for the energy build-up, a Poynting flux of $10^7\,{\rm erg}\,{\rm cm}^{-2}\,{\rm s}^{-1}$ is obtained. Importantly, this is comparable to the input required to balance the coronal losses, lending support to the braiding scenario.

Subsequent studies have estimated the Poynting flux due to photospheric motions using various assumptions about the \emph{average} properties of photospheric flows \citep[][and the MHD models cited above]{vanBallegooijen1986b,Berger1991q,Berger1993b,Zirker1993}. The aim of this paper is to make quantitative estimates of the Poynting flux for a \emph{specific} dataset of observed photospheric velocities. We start from two rigorous lower bounds on $W$ for a given sequence $v_x(x,y,t)$, $v_y(x,y,t)$ derived by \citet{Aly}\footnote{also J.-J. Aly (private communication).}. The first bound (Section \ref{sec:first}) is based, like Parker's estimate, on the net displacement between the two footpoints of each field line. The second bound (Section \ref{sec:top}) is more sophisticated and based on the relative twisting between pairs of field lines. It draws on the well-established idea that entanglement of magnetic field lines puts a lower bound on the energy of a magnetic field \citep{Taylor1974s,Moffatt1985c,Freedman1991}. Our work builds on that of \citet{Berger1993b}, who derived a lower bound for the energy of a braided magnetic field in terms of relative winding between field lines. Here, we remove his assumption that ${\bf B}$ has a uniform vertical component, and modify the bound slightly so that it is computable solely from a sequence of photospheric velocities.

The observational application of the bounds derived in Section \ref{sec:bounds} is presented in Section \ref{sec:results}. It is important to note that, although our lower bounds for $W$ are strict for the chosen Cartesian domain, they depend on the assumed height $h$ of the domain. Moreover, the real coronal magnetic field from a specific photospheric region will likely fill a different shape of domain, and this will also influence the true magnetic energy. However, the lack of definitive methods for coronal magnetic field extrapolation prevents us from accounting for the precise volume. In view of these uncertainties, we find it useful to compare in Section \ref{sec:comps} with two alternative Poynting flux estimates: one obtained with a vector magnetogram, and a hypothetical magnetic field reconstruction having the correct field line connectivity. In addition, to account for possible limitations of our velocity observations, we apply the technique to horizontal velocities taken from a numerical convection simulation. Conclusions are given in Section \ref{sec:conclusions}.

\section{Lower bounds for the Poynting flux} \label{sec:bounds}

\begin{figure}[h]
\begin{center}
\includegraphics[width=4cm]{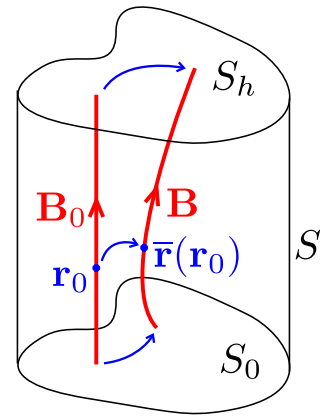}
\caption{Notation for a magnetic field ${\bf B}$ in a cylindrical domain $V$.} \label{fig:defns}
\end{center}
\end{figure}

For the convenience of the reader, this section presents the derivation of the two energy bounds given by \citet{Aly}. To obtain strict lower bounds for the magnetic energy $W$, we consider a magnetic field ${\bf B}(x,y,z)$ in a cylindrical domain $V$ of base $S_0\subset\{(x,y,0)\}$, upper boundary $S_h\subset\{(x,y,h)\}$ and height $h>0$ (Figure \ref{fig:defns}). Denote the vertical surface by $S$. We define $2R_0$ to be the diameter of $S_0$ (i.e., the largest distance between two points on the boundary $\partial S_0$), and $|S_0|$ to be the area of $S_0$. For the applications in this Paper, we will take the cross-section $S_z$ to be rectangular, but the two bounds we derive in this section apply more generally.

Our energy bounds will further assume that ${\bf B}$ is the result of a continuous deformation of the initial uniform field ${\bf B}_0=B_0{\bf e}_z$, with $B_0>0$ constant. This deformation is given by a one-to-one orientation preserving mapping, which we denote $\overline{\bf r}$. It transports a plasma element from ${\bf r}_0=(x_0,y_0,z_0)\in V$ to $\overline{\bf r}({\bf r}_0)=\big(\overline{\bf R}({\bf r}_0),\overline{z}({\bf r}_0)\big) \in V$. Note that capital $\overline{\bf R}$ denotes the $\overline{x}$ and $\overline{y}$ components only. Also, $\overline{\bf r}$ maps each of the boundaries $S_0, S_h, S$ to itself. 
We can write ${\bf B}$ in terms of the initial field $B_0$ and the mapping as
\begin{equation}
{\bf B}\big(\overline{\bf r}({\bf r}_0) \big) = \frac{B_0}{J({\bf r}_0)}\frac{\partial\overline{\bf r}}{\partial z_0}({\bf r}_0).
\label{eqn:bjac}
\end{equation}
Here $J$ is the determinant of $\nabla_0\overline{\bf r}$, and we have used that ${\bf B}$ is a pseudovector ($\nabla_0$ denotes the gradient with respect to the initial point ${\bf r}_0$). From equation \eqref{eqn:bjac}, we can then write the magnetic energy \eqref{eqn:energy} as
\begin{equation}
W = \int_V\frac{B_0^2}{8\pi J}\left|\frac{\partial\overline{\bf r}}{\partial z_0}\right|^2\,d^3x_0.
\end{equation}

Throughout this paper, $W_0$ denotes the energy of the initial uniform field,
\begin{equation}
W_0 = \frac{B_0^2|V|}{8\pi},
\end{equation}
where $|V|$ is the volume of $V$.

\subsection{First lower bound} \label{sec:first}

A first lower bound for the free energy may be derived using only the footpoint connectivity of field lines \citep{Aly}. In particular, if ${\bf R}_0$ is the footpoint of a given field line in ${\bf B}_0$ (the footpoints on both $S_0$ and $S_h$ are the same since the field line is vertical), then let $\boldsymbol\xi({\bf R}_0)$ denote the horizontal distance between the corresponding footpoints after the deformation:
\begin{align}
\boldsymbol\xi({\bf R}_0) &= \overline{\bf R}({\bf R}_0,h) - \overline{\bf R}({\bf R}_0,0),\\
&= \overline{\bf r}({\bf R}_0,h) - h{\bf e}_z - \overline{\bf r}({\bf R}_0,0),\\
&= \int_0^h\frac{\partial\overline{\bf r}}{\partial z_0}({\bf R}_0,z_0)\,dz_0 - h{\bf e}_z. \label{eqn:xir}
\end{align}
The quantity we wish to bound is
\begin{align}
\frac{W}{W_0} &= \frac{1}{|V|}\int_V\frac{1}{J}\left|\frac{\partial\overline{\bf r}}{\partial z_0}\right|^2\,d^3x_0 = \nonumber\\
&\qquad\frac{1}{|V|^2}\left(\int_V\frac{1}{J}\left|\frac{\partial\overline{\bf r}}{\partial z_0}\right|^2\,d^3x_0\right)\left(\int_VJ\,d^3x_0\right).
\end{align}
Applying the Cauchy-Schwarz inequality gives
\begin{equation}
\frac{W}{W_0} \geq \frac{1}{|V|^2}\left(\int_V\left|\frac{\partial\overline{\bf r}}{\partial z_0} \right|\,d^3x_0\right)^2 \geq \frac{1}{|V|^2}\left(\int_{S_0}\left|\int_0^h\frac{\partial\overline{\bf r}}{\partial z_0}\,dz_0 \right| \,d^2x_0 \right)^2,
\end{equation}
in which substitution of \eqref{eqn:xir} leads to the bound
\begin{align}
\frac{W}{W_0} &\geq \left[\int_{S_0}\left(1 + \left|\frac{\boldsymbol\xi}{h}\right|^2 \right)^{1/2}\frac{d^2x_0}{|S_0|} \right]^2.
\label{eqn:bound1}
\end{align}
This bound for the free energy depends only on the horizontal distances between end-points of each field line, as well as the height $h$ of the domain.

\subsection{Second lower bound} \label{sec:top}

\citet{Aly} derived a second lower bound for $W/W_0$ in terms of the relative twist of pairs of field lines. This is a generalisation of a bound derived by \citet{Berger1993b} under the much more restrictive condition that $B_z=B_0$ for all time. For completeness we present here the derivation of the new bound, where we shall assume only that $B_z > 0$ everywhere in $V$. It will be convenient to decompose ${\bf B}$ into horizontal and vertical components ${\bf B}={\bf b}_\perp + B_z{\bf e}_z$, and to further decompose $B_z=B_0 + b_z$, where the constant $B_0$ is the initial field. 

The \emph{relative twist} $\phi({\bf R}^z,\widetilde{\bf R}^z)$ of two field lines ${\bf R}^z$, $\widetilde{\bf R}^z$, rooted at two points ${\bf R}^0$, $\widetilde{\bf R}^0$ on $S_0$, is defined as the net angle swept out by the vector  ${\bf X}^z = \widetilde{\bf R}^z-{\bf R}^z$ as $z$ goes from $0$ to $h$. This may be expressed as
\begin{equation}
\phi({\bf R}^z,\widetilde{\bf R}^z) = \int_0^h\frac{\partial\varphi^z}{\partial z}({\bf R}^z,\widetilde{\bf R}^z)\,dz,
\label{eqn:phi}
\end{equation}
where
\begin{equation}
\tan\varphi^z = \frac{{\bf X}^z\cdot{\bf e}_y}{{\bf X}^z\cdot{\bf e}_x}.
\end{equation}
Since ${\bf R}^z$ and $\widetilde{\bf R}^z$ are magnetic field lines, we can write (after some algebra)
\begin{equation}
\frac{\partial\varphi^z}{\partial z}({\bf R}^z,\widetilde{\bf R}^z) = \left[\left(\frac{{\bf b}_\perp}{B_z}\right)(\widetilde{\bf R}^z,z) -  \left(\frac{{\bf b}_\perp}{B_z}\right)({\bf R}^z,z) \right]\cdot\left(\frac{\widehat{\bf u}}{X^z} \right),
\label{eqn:dphidz}
\end{equation}
where $\widehat{\bf u}={\bf e}_z\times{\bf X}^z/X^z$ \citep[e.g.,][]{Berger1993b}. For any pair of field lines, $\phi({\bf R}^z,\widetilde{\bf R}^z)$ is invariant under any ideal deformation that fixes the end-points: it is a topological quantity. Our energy bound will be expressed in terms of the average twist over all pairs of field lines, namely
\begin{equation}
w^* = \frac{1}{16\pi R_0}\int_{S_0\times S_0}\big|\phi({\bf R}^z,\widetilde{\bf R}^z)\big|B_z({\bf R}^0,0)B_z(\widetilde{\bf R}^0,0)\,d^2x\,d^2\widetilde{x}.
\label{eqn:wstar}
\end{equation}
This differs slightly from \citet{Berger1993b} who took the absolute value inside the $\phi$-integral \eqref{eqn:phi}. (In that case $\phi$ is no longer a topological invariant, and the bound is more restrictive, being geometrical rather than topological in character.) With no absolute value sign, \eqref{eqn:wstar} would give the relative magnetic helicity, for an appropriate reference field \citep{Berger1986}. The quantity $w^*$ can be measured from photospheric observations if one knows the initial photospheric distribution of $B_z$ and the subsequent pattern of footpoint motions: this is the basis of our computations in Section \ref{sec:results}.

To bound $W/W_0$ in terms of $w^*$, we start by substituting \eqref{eqn:phi} and \eqref{eqn:dphidz} into \eqref{eqn:wstar}. Applying the triangle inequality and relabelling one of the terms, we get
\begin{align}
w^* &\leq \frac{1}{8\pi R_0}\int_{S_0\times S_0}\left(\int_0^h\left|\left(\frac{{\bf b}_\perp}{B_z} \right)({\bf R}^z,z)\cdot \left(\frac{\widehat{\bf u}}{X^z} \right) \right|\,dz \right)
\nonumber\\
&\qquad\qquad\qquad\qquad\qquad \cdot
B_z({\bf R}^0,0)B_z(\widetilde{\bf R}^0,0)\,d^2x\,d^2\widetilde{x}.
\end{align}
Writing the horizontal integrals over $S_z$ instead of $S_0$ introduces a factor $B_z({\bf R}^z,z)/B_z({\bf R}^0,0)$ for each integral, giving
\begin{equation}
w^* \leq \frac{1}{8\pi R_0}\int_V b_\perp\left(\int_{S_z}\frac{|\widehat{\bf b}_\perp\cdot\widehat{\bf u}|}{X^z}B_z(\widetilde{\bf R}^z,z) d^2\widetilde{x}\right)\,d^3x,
\end{equation}
where $\widehat{\bf b}_\perp = {\bf b}_\perp({\bf R}^z,z)/|{\bf b}_\perp({\bf R}^z,z)|$. Since $w^*\geq 0$ it follows that $|w^*|=w^*$, and the triangle inequality gives
\begin{align}
8\pi R_0 |w^*| &\leq B_0\int_V b_\perp\left(\int_{S_z}\frac{|\widehat{\bf b}_\perp\cdot\widehat{\bf u}|}{X^z} d^2\widetilde{x}\right)\,d^3x \nonumber\\
&\qquad\qquad + \int_V b_\perp \left(\int_{S_z}\frac{|\widetilde{b_z}|}{X^z}\,d^2\widetilde{x}\right)\,d^3x,
\label{eqn:ineq1}
\end{align}
where $\widetilde{b_z}=b_z(\widetilde{\bf R}^z,z)$. Our task is then to bound the two integrals in \eqref{eqn:ineq1} in terms of the energy $W/W_0$.

The first integral in \eqref{eqn:ineq1} -- which we denote $I_1$ -- is similar to equation (10) of \citet{Berger1993b} and may be bounded in a similar way. In particular, let $I({\bf R}^z) = (1/R_0)\int_{S_z}|\widehat{\bf b}_\perp\cdot\widehat{\bf u}|/X^z\,d\widetilde{x}$. Then the maximum value of $I({\bf R}^z)$ over all possible $\widehat{\bf b}_\perp$, which we denote $m({\bf R}^z)$, is a function depending only on the geometry of $S_0$. For example, if $S_0$ is a disk, then one can show (by explicit but tedious calculation) that
\begin{equation}
m({\bf R}^z) = 2\sqrt{1 - \left(\frac{R^z}{R_0}\right)^2} + 2\frac{R_0}{R^z}\arcsin\left(\frac{R^z}{R_0}\right).
\end{equation}
(Note that an incorrect expression is given in \citealp{Berger1993b}.) We then have
\begin{equation}
I_1 \leq  R_0B_0\int_V b_\perp m({\bf R}^z)\,d^3x.
\label{eqn:i1}
\end{equation}
Defining $\mu^2 = |S_0|^{-1}\int_{S_z}m^2\,d^2x$, and applying the Cauchy-Schwarz inequality, we obtain
\begin{equation}
I_1 \leq \sqrt{|V|}R_0B_0\mu\left(\int_V b_\perp^2 d^3x \right)^{1/2}.
\label{eqn:boundI1}
\end{equation}
A numerical integration shows that (for the disk) $\mu^2\approx 13.137$.

Now we look for a bound of the second integral in \eqref{eqn:ineq1}, which we denote $I_2$. The Cauchy-Schwarz inequality yields
\begin{equation}
I_2 \leq \left(\int_V b_\perp^2\,d^3x \right)^{1/2}\left(\int_V \left[\int_{S_z}\frac{|\widetilde{b}_z|}{X^z}\,d^2\widetilde{x} \right]^2\,d^3x \right)^{1/2}.
\end{equation}
Applying Cauchy-Schwarz again to the $\widetilde{x}$-integral, we obtain
\begin{equation}
I_2 \leq R_0\nu\left(\int_V b_\perp^2\,d^3x \right)^{1/2}\left(\int_V b_z^2\,d^3x \right)^{1/2},
\label{eqn:boundI2}
\end{equation}
where $\nu^2 = R_0^{-2}\int_{S_z\times S_z}(X^z)^{-2}\,d^2\widetilde{x}\,d^2x$ is another geometric constant. For any domain $S_0$, we have $\nu \leq 2\pi$ (with equality when $S_0$ is a disk).

We may now use \eqref{eqn:boundI1} and \eqref{eqn:boundI2} in \eqref{eqn:ineq1} to find
\begin{equation}
(8\pi R_0)^2(w^*)^2 \leq R_0^2\left(\int_V b_\perp^2\,d^3x \right)\left[\sqrt{|V|}B_0\mu + \nu\left(\int_V b_z^2\,d^3x \right)^{1/2} \right]^2
\end{equation}
This may be re-arranged to yield the inequality
\begin{equation}
\frac{W}{W_0} \geq 1 + s + \frac{(w^*/W_0)^2}{(\mu + \nu\sqrt{s})^2},
\label{eqn:bound2a}
\end{equation}
where $s = (1/W_0)\int_V b_z^2/(8\pi)\,d^3x$. Under the additional assumption that $b_z=0$, inequality \eqref{eqn:bound2a} would reduce to
\begin{equation}
\frac{W}{W_0} \geq 1 + \frac{1}{\mu^2}\left(\frac{w^*}{W_0}\right)^2,
\label{eqn:bergerbound}
\end{equation}
which is equivalent to the bound of \citet{Berger1993b}. In general, we allow for any distribution of $b_z$, and our bound is
\begin{equation}
\frac{W}{W_0} \geq \min_{s\geq 0}\left\{1 + s + \frac{(w^*/W_0)^2}{(\mu + \nu\sqrt{s})^2} \right\}.
\label{eqn:bound2}
\end{equation}
If $w^*=0$ then the minimum occurs at $s=0$. Otherwise, let $x=\sqrt{s}$ so that our function becomes $g(x) = 1 + x^2 + a^2/(b + cx)^2$, where $b, c > 0$. A minimum must satisfy
\begin{equation}
c^3x^4 + 3bc^2x^3 + 3b^2cx^2 + b^3x - a^2c = 0,
\end{equation}
and applying Descartes' rule of signs shows that there is exactly one root for positive $x$ (and hence $s$). This root gives the minimum value in \eqref{eqn:bound2}.

\subsection{Simple example} \label{sec:test}

\begin{figure}[h]
\begin{center}
\includegraphics[width=0.45\columnwidth]{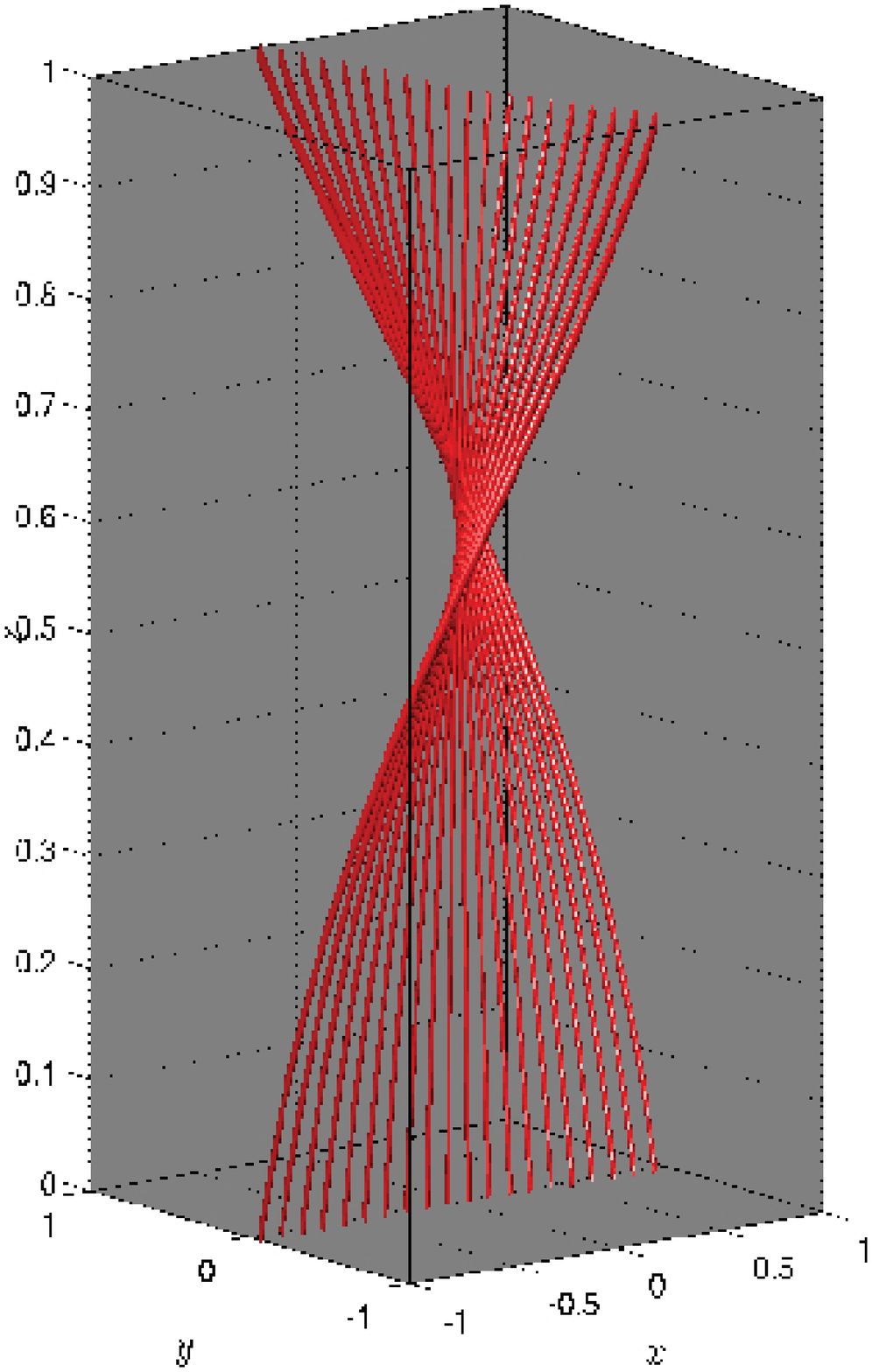}
\includegraphics[width=0.45\columnwidth]{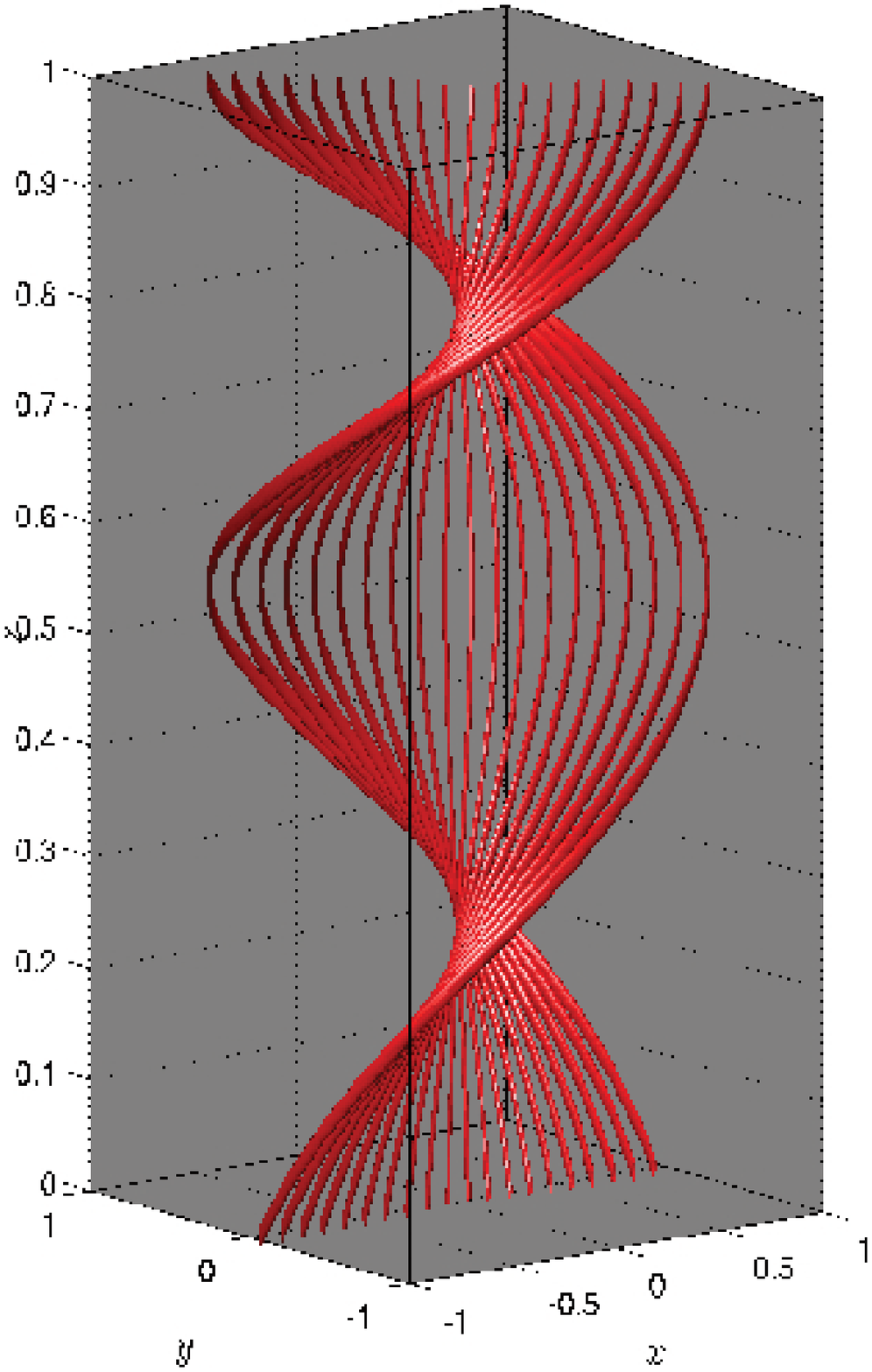}
\includegraphics[width=\columnwidth]{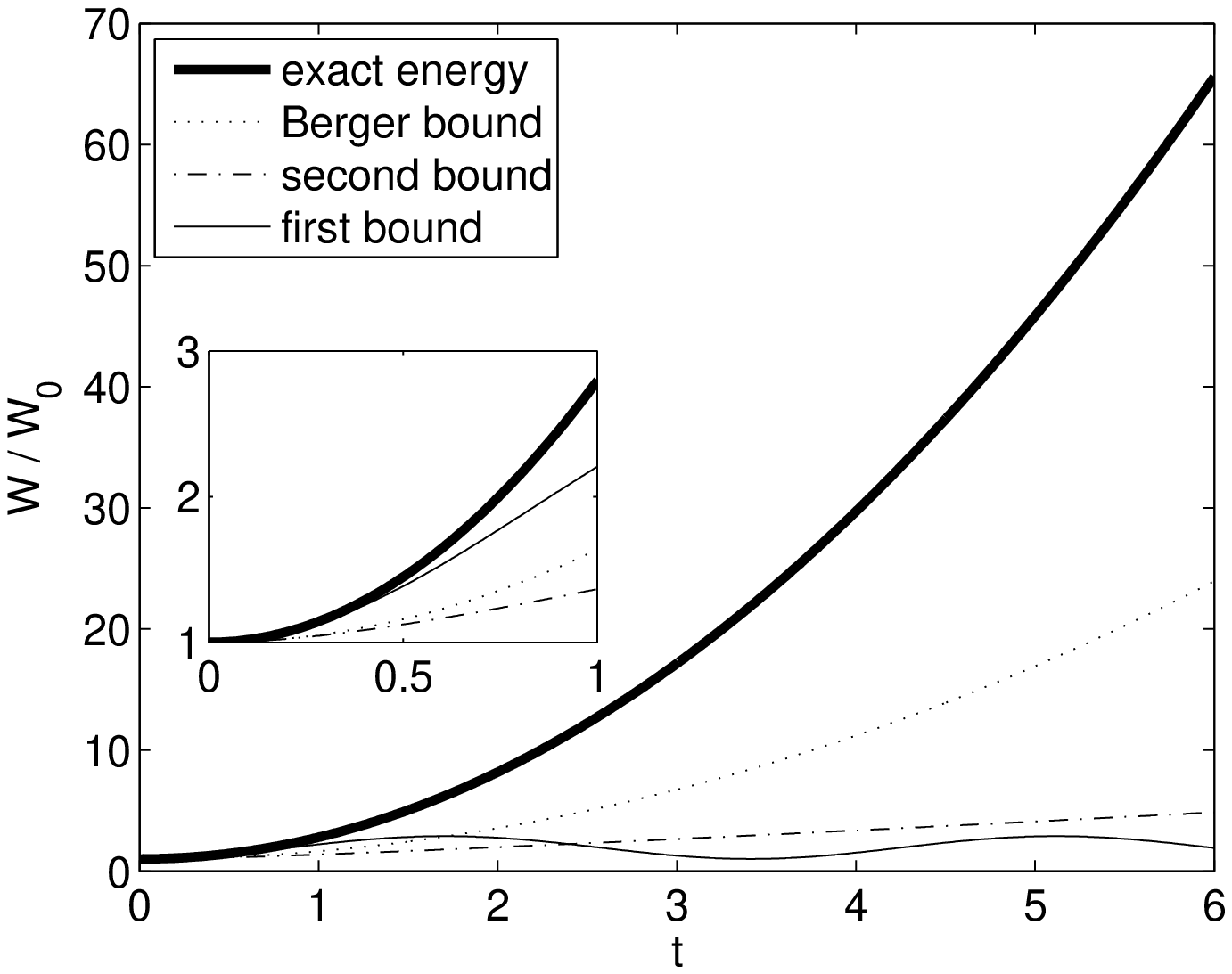}
\caption{Comparison of the exact energy $W/W_0$ and three different bounds for the analytical magnetic field in Section \ref{sec:test}, with $h=1$, $R_0=1$, $v_0=1$. The two upper panels show field lines at $t=1$ and $t=3$ respectively.} \label{fig:test}
\end{center}
\end{figure}

To compare the two lower bounds \eqref{eqn:bound1} and \eqref{eqn:bound2}, it is instructive to look at a simple example before considering the observed data. We take $S_0$ to be a disk, and apply a large-scale twist to the initial field ${\bf B}_0=B_0{\bf e}_z$ by solving the ideal MHD induction equation with the specified velocity ${\bf v}=v_0rf(z){\bf e}_\phi$. Here $(r,\phi,z)$ are standard cylindrical coordinates. The resulting magnetic field is ${\bf B}(r,z,t)=v_0B_0rf'(z)t{\bf e}_\phi + B_0{\bf e}_z$. For illustration we choose $f(z)=z^2 + \sin z$ (Figure \ref{fig:test}). The exact magnetic energy of this field is
\begin{equation}
\frac{W}{W_0} = 1 + \frac{v_0^2R_0^2t^2}{2h}\left(\frac{4h^3}{3} + 4h\sin h + 4\cos h -4 + \frac{h}{2} + \frac{\sin h \cos h}{2} \right),
\end{equation}
which increases quadratically with time. The first lower bound \eqref{eqn:bound1} may be computed explicitly, giving
\begin{equation}
\frac{W}{W_0} \geq \left(\frac{h^2}{R_0^2 G^2}\left[\frac{2}{3}\left(1 + \frac{G^2R_0^2}{h^2} \right)^{3/2} - \frac{2}{3}  \right] \right)^2,
\end{equation}
where $G(t) = 2\sin\big[(f(h)-f(0))v_0t/2\big]$. Notice that $G$, and hence the bound, is a periodic function of $t$ (as seen in Figure \ref{fig:test}). This reflects the limitation that the bound cannot detect net rotations of field lines, as it depends only on the end-points. 

For the second lower bound \eqref{eqn:bound2}, we find explicitly that
\begin{equation}
\frac{w^*}{W_0} = \frac{\pi R_0|(f(h) - f(0))v_0t|}{2h},
\label{eqn:bound2test}
\end{equation}
and the bound \eqref{eqn:bound2} is readily evaluated numerically. Comparing this with the first bound (Figure \ref{fig:test}), we see that \eqref{eqn:bound2test} is not periodic, and increases monotonically in time as the field lines become more and more twisted. For $t>2.5$, this gives a better estimate of the true energy than the first bound, although it is not tight. In fact, Berger's bound \eqref{eqn:bergerbound} gives a much better estimate for $t>0.5$, since the true field maintains constant $B_z=B_0$. This additional information will not be available from the observational data, so we need to allow for the field reaching a lower energy by adopting varying $B_z$ in the corona. We remark that this is a rather extreme example, with a large-scale twist filling the whole of $V$, and that for small $t$ the first bound \eqref{eqn:bound2} is actually tightest, as seen from the inset in Figure \ref{fig:test}. We also find the first bound to be tightest in the observed data, described in the next section.

\section{Observational results} \label{sec:results}

\subsection{Data reduction}

\begin{figure}[h]
\begin{center}
\includegraphics[width=\columnwidth]{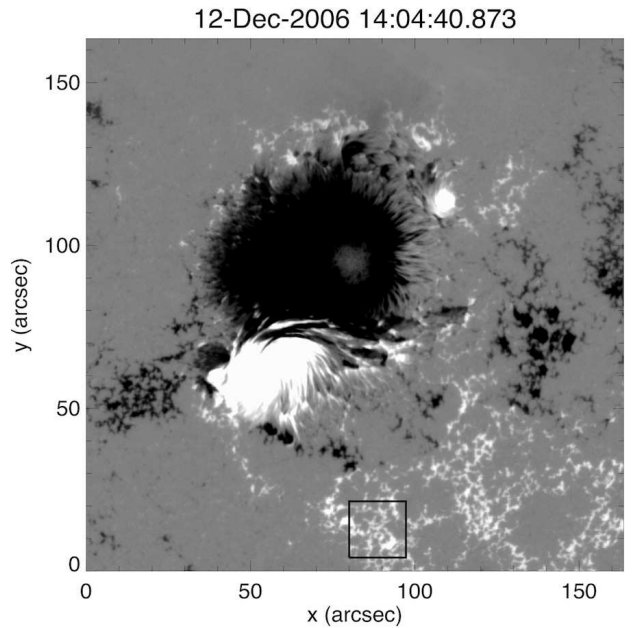}
\caption{Line-of-sight magnetogram showing the location of the observed plage region (black box) in the Hinode/NFI field of view, away from the main sunspots of active region 10930.} \label{fig:region}
\end{center}
\end{figure}

We use the same velocity data as \citet{Yeates2012}, derived by local correlation tracking in line-of-sight magnetograms. As in that paper, we focus on a unipolar plage region of size  $12.4\,\textrm{Mm}\times 12.4\,\textrm{Mm}$ (approximately $17''\times 17''$), near active region 10930 but away from the sunspots (Figure \ref{fig:region}). The original Stokes $V/I$ maps were taken by Hinode/NFI (Narrowband Filter Imager) observations in Fe I $6302\AA$ \citep{Tsuneta2008}, and the sequence runs from 14:00UT on 12 December to 02:58UT on 13 December 2006, at a cadence of $\sim 121\,\textrm{s}$. The line-of-sight magnetic field strength was empirically calibrated \citep[following][]{isobe2007}, and the velocity field was extracted from the magnetograms using Fourier local correlation tracking \citep[FLCT;][]{welsch2004,fisher2008}. Full details of the procedure, including its optimisation, are given in \citet{Welsch2012d}. As in \citet{Yeates2012}, high-frequency noise in the velocity field was removed with minimal disturbance to the well-resolved regions by applying a low-pass spatial filter in Fourier space. The result is a time-sequence of horizontal photospheric velocities $v_x$, $v_y$, at horizontal resolution of approximately $230\,{\rm km}$. Note that the mean flow speed in these observations is of the order $0.1\,\textrm{km}\,\textrm{s}^{-1}$, rather slower than reported speeds for granular flows \citep[$\sim 1\,\textrm{km}\,\textrm{s}^{-1}$,][]{rieutord2010}. This underestimate may be a combined effect of the observational resolution and the correlation tracking \citep{Welsch2007}. However, it may also reflect either slower motion of magnetic features compared to the underlying plasma velocity, or inhibition of the convective flow by strong magnetic fields in our plage region. 

\subsection{Lower bounds on the Poynting flux} \label{sec:obsbound}

We now use \eqref{eqn:bound1} and \eqref{eqn:bound2} to put lower bounds on the Poynting flux resulting from our observed sequence of photospheric velocities. Recall that \eqref{eqn:bound1} and \eqref{eqn:bound2} bound the magnetic energy $W/W_0$ of a final magnetic field ${\bf B}(x,y,z)$ that has been generated by ideal deformation of an initially uniform field $B_0{\bf e}_z$. But we do not need to know the precise deformation: in fact we need only the end-point connectivity of field lines for \eqref{eqn:bound1}, and the relative twist of each pair of field lines for \eqref{eqn:bound2}. Both of these quantities can be computed by knowing only the sequence of footpoint motions on $S_0$ and on $S_h$. In this Paper, we shall assume that the footpoints on $S_0$ remain fixed, while those on $S_h$ move according to the observed time-sequence of $v_x$, $v_y$. (On average, this likely means that we will underestimate the Poynting flux compared to the real corona where field lines are undergoing largely uncorrelated footpoint motions at either end.)

To match the observed region, it is convenient to take $S_0$ (and hence every cross-section $S_z$) to be a square of side $L$, rather than a disk. Hence $R_0=L/\sqrt{2}$. For a square, the geometric constants $\mu$, $\nu$ used in the second lower bound (Section \ref{sec:top}) may be taken to be $\mu^2=14.146$ and $\nu=2\pi$. Both are upper bounds -- the former uses the fact that $m({\bf R}^z)$ in \eqref{eqn:i1} is bounded above by its value for a disk of radius $L/\sqrt{2}$ ($m^2$ is then integrated over the square). The height $h$ of the domain is not constrained by the observations -- we shall consider its effect below.

\begin{figure}[h]
\begin{center}
\includegraphics[width=\columnwidth]{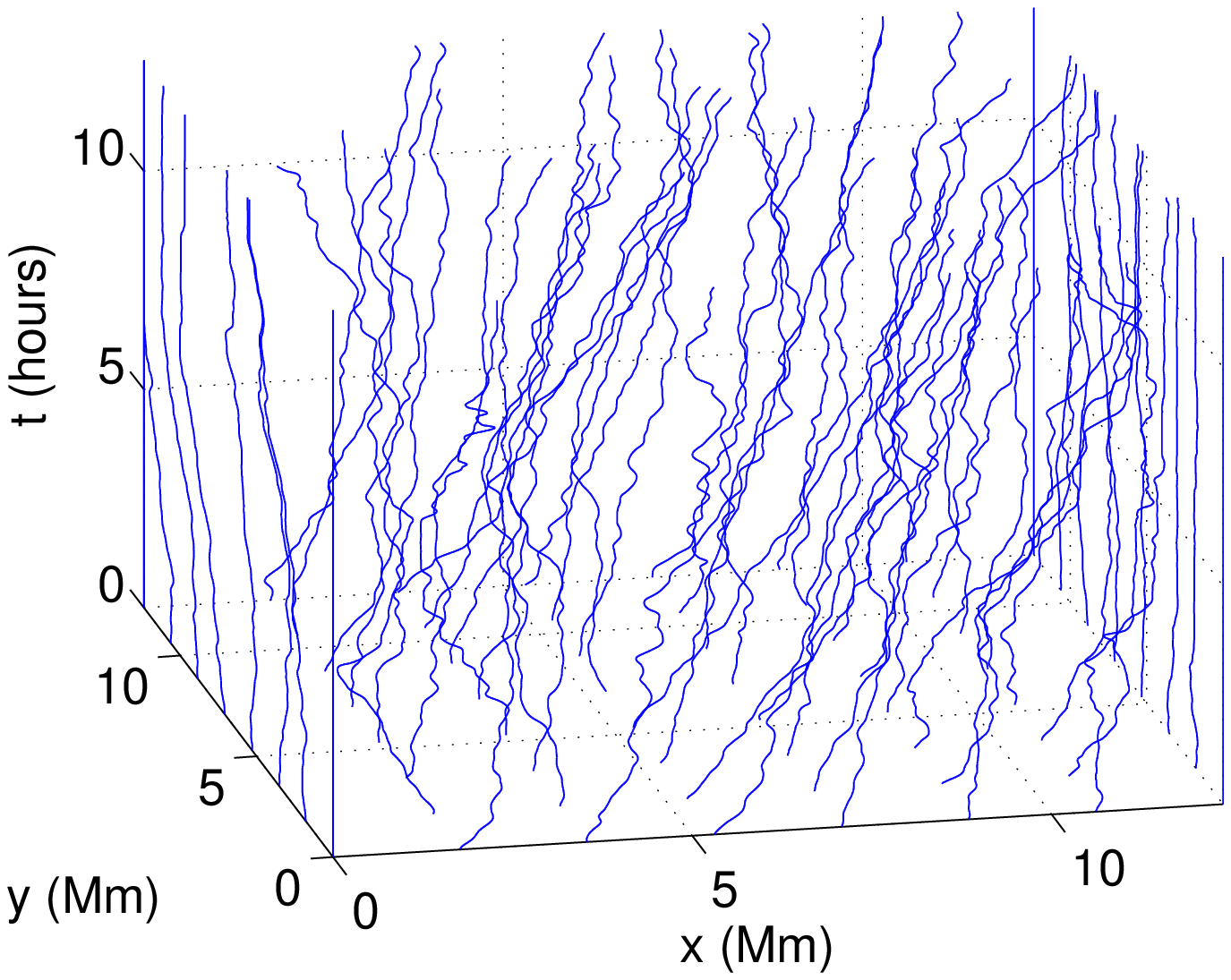}
\includegraphics[width=\columnwidth]{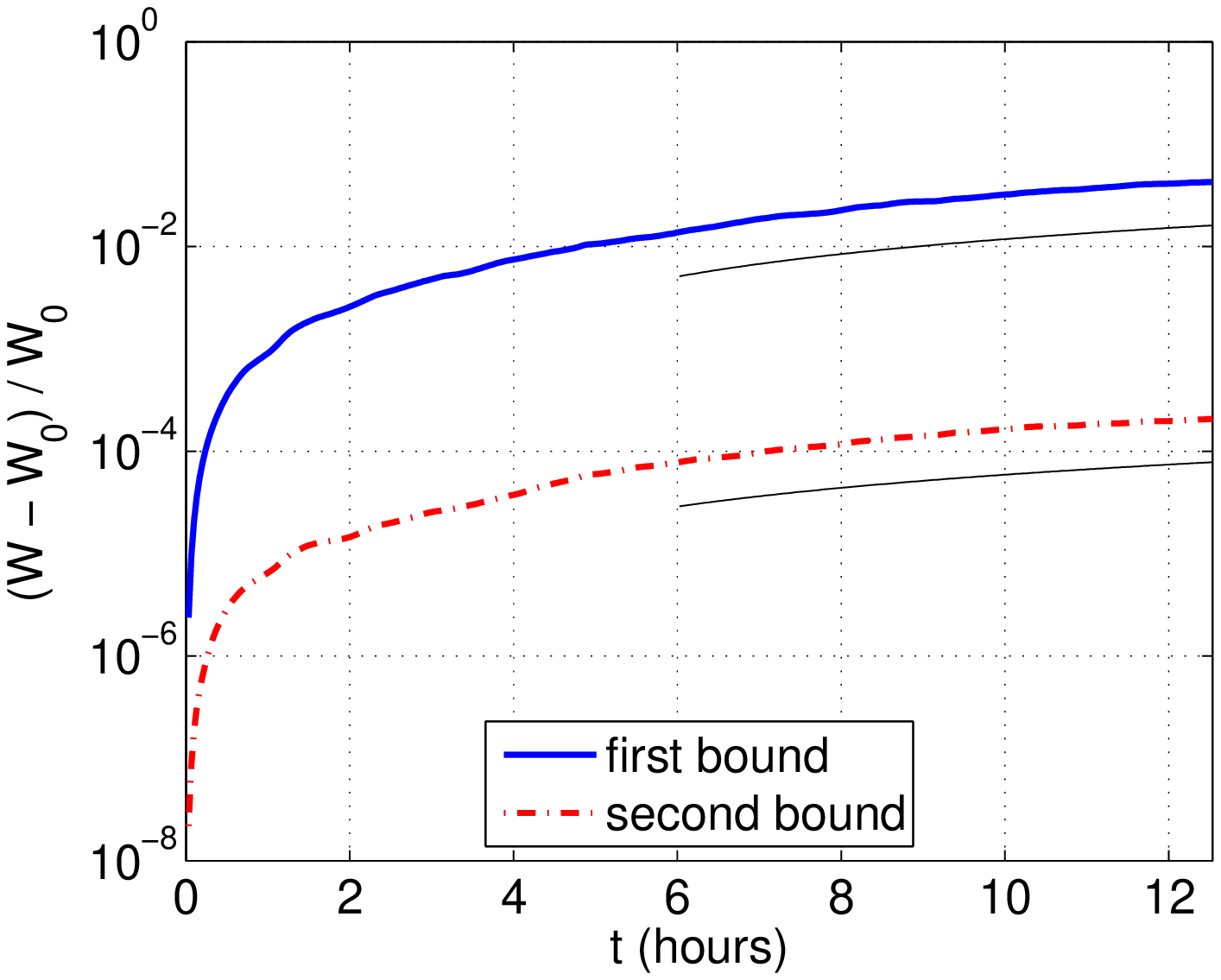}
\caption{Trajectories of the observed velocity field (upper panel) and the two lower bounds calculated for a similar set of 900 trajectories (lower panel). The vertical axis shows $(W-W_0)/W_0$ on a log-scale. Thin lines show the slope of linear fits for $t > 6\,{\rm hr}$.} \label{fig:welsch}
\end{center}
\end{figure}

Under these assumptions, we have computed the bounds \eqref{eqn:bound1} and \eqref{eqn:bound2} numerically by integrating trajectories of the observed velocity field (Figure \ref{fig:welsch}). This allows us to compute both $\boldsymbol\xi$ and $\phi({\bf R}^z,\widetilde{\bf R}^z)$, as required for the bounds. We also need $B_z$ on $S_0$ to compute $w^*$, but we simply have $B_z|_{S_0}=B_0$. There is one complication: our bounds assume that field lines are confined within $V$ at all times. To prevent violation of this constraint on the side boundary, we artificially set the normal velocity to  zero on the side boundaries so that trajectories do not leave (or enter) $V$. This likely has the effect of reducing the estimated energy a little.

The resulting bounds on the free energy are shown in Figure \ref{fig:welsch} as a function of time. The quantity plotted is $(W-W_0)/W_0$, so as to facilitate showing both bounds on the same logarithmic scale. The most noticeable result is that the first bound is approximately 100 times larger than the second bound. This reflects the fact that most trajectories seem to have small relative twist.

\begin{figure}[h]
\begin{center}
\includegraphics[width=\columnwidth]{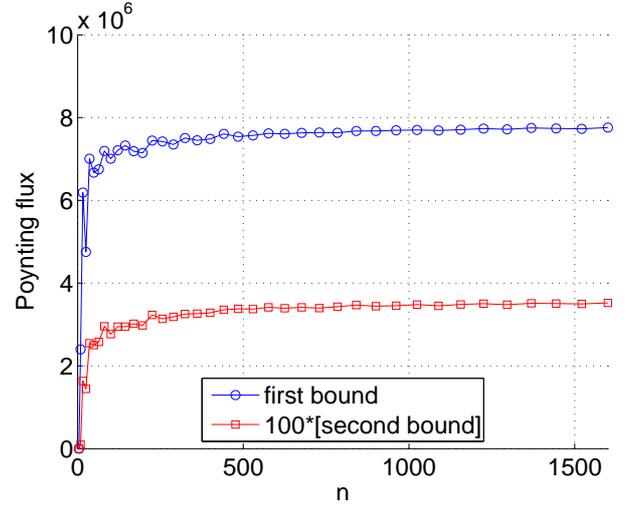}
\caption{Convergence of the numerical estimates of the bounds as the number $n$ of field lines in the calculation is increased. The units of Poynting flux (per unit area) are ${\rm erg}\,{\rm cm}^{-2}\,{\rm s}^{-1}$, and we have taken $B_0=350\,{\rm G}$, $h=L$.} \label{fig:welsch_convergence}
\end{center}
\end{figure}

To calculate the Poynting flux implied by our energy bounds, we notice that, from about $t=6\,{\rm hr}$ onwards, $W/W_0$ grows approximately linearly in time. A linear fit gives this linear growth rate $s$, which then gives the Poynting flux
\begin{equation}
\frac{dW}{dt} = W_0s = \frac{L^2hB_0^2s}{8\pi}
\end{equation}
for our assumed ideal evolution. A numerical estimate requires values for $L$, $h$ and $B_0$. Here $L\approx 12\,{\rm Mm}$ is fixed by the region of observation, and we take $B_0=350\,{\rm G}$ (close to the observed mean value of line-of-sight magnetic field in the region). Figure \ref{fig:welsch_convergence} shows the two estimates expressed as Poynting flux per unit area, $L^{-2}dW/dt$, and verifies that the estimates converge as more and more trajectories are used for the calculation with $h=L$.

\begin{figure}[h]
\begin{center}
\includegraphics[width=\columnwidth]{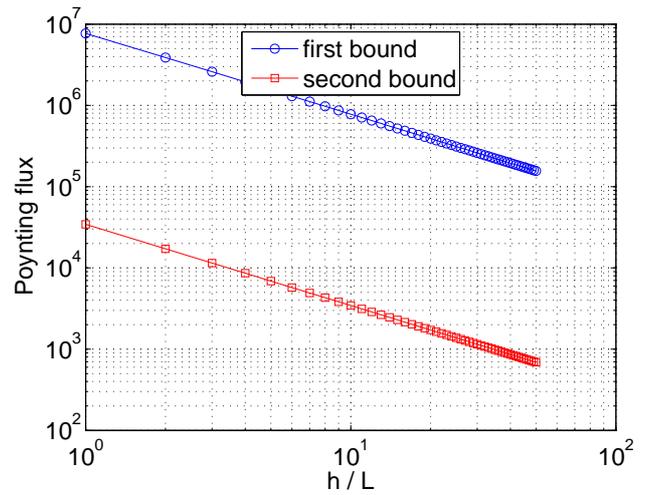}
\caption{Dependence of each of the bounds on domain size $h$, shown on a log-log scale. The units of Poynting flux (per unit area) are ${\rm erg}\,{\rm cm}^{-2}\,{\rm s}^{-1}$, and we have taken $B_0=350\,{\rm G}$. Both lines have slope $-1$.} \label{fig:welsch_h}
\end{center}
\end{figure}

Although the Poynting flux with $h=L$ for the first bound is approximately $8\times 10^7\,{\rm erg}\,{\rm cm}^{-2}\,{\rm s}^{-1}$, close to the required rate to heat the corona, it should be noted that this estimate goes down if the height $h$ of the domain is increased (think of this as the length of the coronal loop). In fact, the Poynting fluxes from both bounds decrease like $h^{-1}$ (as did Parker's original simple estimate). This is shown in Figure \ref{fig:welsch_h}. For a more realistic domain height of $h=8L(\approx 100\,{\rm Mm})$, the largest lower bound gives only $\approx 10^6\,{\rm erg}\,{\rm cm}^{-2}\,{\rm s}^{-1}$. To understand why the energy bounds scale like $h^{-1}$, consider a magnetic field ${\bf B}={\bf B}_\perp(x,y,z) + B_z(x,y,z){\bf e}_z$ in a domain $0<z<1$, where ${\bf B}_\perp$ denotes the horizontal components. This has energy
\begin{equation}
W = \frac{1}{8\pi}\int_{S_0}\int_0^1\big(B_\perp^2(x,y,z) + B_z^2(x,y,z)\big)\,dzdxdy.
\end{equation}
The stretched field ${\bf B}'=h^{-1}{\bf B}_\perp(x,y,z/h) + B_z(x,y,z/h){\bf e}_z$ has the same field line connectivity in the domain $0<z<h$ \citep{Aly2010i}, and the new energy is
\begin{align}
W' &= \frac{1}{8\pi}\int_{S_0}\int_0^h\left(\frac{1}{h^2}B_\perp^2(x,y,z'/h) + B_z^2(x,y,z'/h)\right)\,dz'dxdy,\\
&= \frac{1}{8\pi}\int_{S_0}\int_0^1\left(\frac{1}{h}B_\perp^2(x,y,z) + hB_z^2(x,y,z)\right)\,dzdxdy.
\end{align}
Now if $B_z\approx B_0$ (i.e., the field has a strong vertical component), then the second term in the integrand is essentially $W_0$ for the new field. So $W-W_0$ is essentially the $B_\perp$ term, leading to $dW/dt \sim h^{-1}$. Thus the scaling with $h^{-1}$ is not an inaccuracy of our energy bounds, but rather arises because it is possible to reduce the free energy of a magnetic field by stretching the domain. 

\section{Comparisons} \label{sec:comps}

To get some idea of the tightness of our lower bounds on the Poynting flux, we now consider three independent energy estimates. The first is a direct calculation of the Poynting flux from a vector magnetogram (Section \ref{sec:mag}) and the second is a reconstructed magnetic field with the correct field-line connectivity (Section \ref{sec:recon}). The third (Section \ref{sec:sim}) is not based on the observations at all, but repeats the calculation for a sequence of velocities from a numerical convection simulation.

\subsection{Vector magnetogram} \label{sec:mag}

\begin{figure}[h]
\begin{center}
\includegraphics[width=\columnwidth]{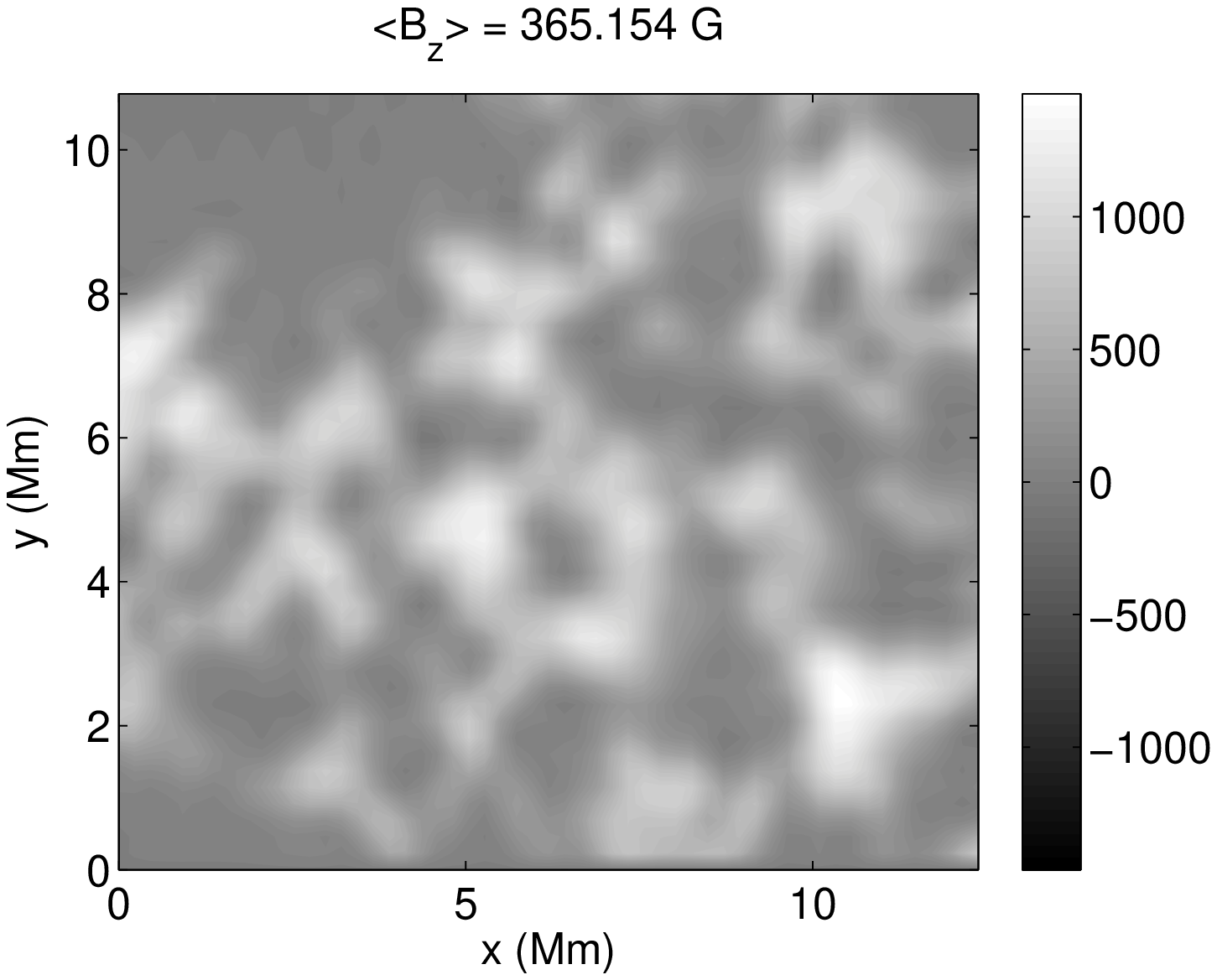}
\includegraphics[width=\columnwidth]{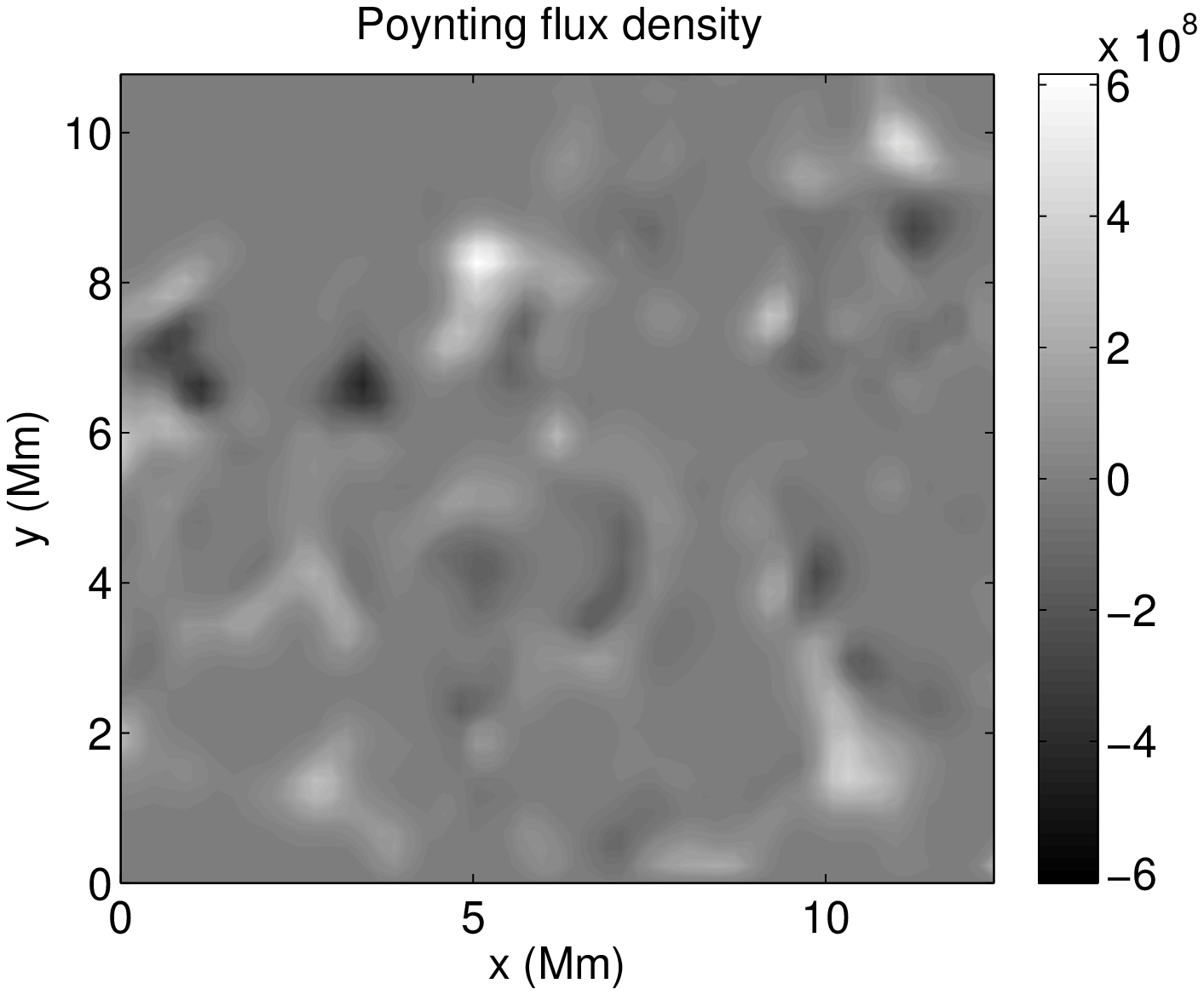}
\caption{Vector magnetogram measurements at 21:04 on 2006 December 12. The upper panel shows $B_z$ from Hinode/SP and the lower panel shows the Poynting-flux density $-B_z(v_xB_x + v_yB_y)/(4\pi)$ using the Hinode/SP $B_x$, $B_y$ and the velocity observations from Section \ref{sec:results}.} \label{fig:welsch_magnetic}
\end{center}
\end{figure}

By combining a vector magnetogram with our measurements of $v_x$ and $v_y$, we may evaluate the last term in equation \eqref{eqn:poyntingS0} directly, subject to combined uncertainties in the velocity and magnetogram measurements (notably the $180^\circ$ ambiguity). This term represents the dominant contribution to the photospheric Poynting flux when ${\bf B}$ is primarily vertical, as we expect in our unipolar plage region. To obtain $B_x$ and $B_y$, we extracted a sub-region from a vector magnetogram analyzed by \citet{Schrijver2008}, taken around 21:04 on 2006 December 12 by Hinode/SP \citep[Spectro-Polarimeter;][]{Tsuneta2008}. The data were co-registered with the NFI line-of-sight magnetic field from which $v_x$ and $v_y$ were extracted, then interpolated to the same resolution. The vertical magnetic field $B_z$ from SP, and the computed Poynting flux density are shown in Figure \ref{fig:welsch_magnetic}(b). Although there are regions of both positive and negative Poynting flux density, the (signed) average Poynting flux per unit area in this region is $1.67\times 10^7\,{\rm erg}\,{\rm cm}^{-2}\,{\rm s}^{-1}$.

\subsection{Topological field reconstruction} \label{sec:recon}

Another estimate of the energy can be obtained by taking the observed trajectories (Figure \ref{fig:welsch}) and assuming them to be the field lines of a magnetic field. Given $B_z$ on one boundary, this uniquely defines a magnetic field with the correct field-line connectivity \citep{Yeates2012}, whose energy we can calculate for a given choice of $h$. In contrast with our lower bounds, this is likely to be an over-estimate of the true energy in the coronal field, since the trajectories are not very smooth: the real magnetic field might be expected to relax to remove such fluctuations on the faster Alfv\'en timescale. (On the other hand, we are neglecting the additional energy that may result from motions at the opposite footpoints.)

To define the magnetic field, let
\begin{equation}
{\bf B}^T(x,y,z)=\lambda(x,y,z)\big(v_x(x,y,z){\bf e}_x + v_y(x,y,z){\bf e}_y + {\bf e}_z \big),
\end{equation}
where ${\bf v}(x,y,t)$ is the observed photospheric velocity sequence. In this way, the $z$ coordinate in the resulting magnetic field corresponds to time in the observed velocity field, and the magnetic field lines have the same topology as the footpoint trajectories. The function $\lambda(x,y,z)$ is uniquely determined by the condition that $\nabla\cdot{\bf B}=0$ along with its distribution $\lambda(x,y,0)=B_0$ on $S_0$ \citep[see][]{Yeates2012}. However, the height $h$ of the domain is then equal to $T$ (the maximum time in the observations). We can stretch the magnetic field to a domain of height $h$ while preserving the field-line connectivity using the same idea as in Section \ref{sec:obsbound}. In particular, define $z'=zh/T$ and
\begin{align}
{\bf B}^h(x,y,z') &= \frac{T}{h}\Big[B^T_x(x,y,z'T/h){\bf e}_x
 + B^T_y(x,y,z'T/h){\bf e}_y\Big] \nonumber\\
&\hspace{3.5cm} + \quad B^T_z(x,y,z'T/h){\bf e}_z.
\end{align}
In terms of ${\bf B}^T$, the energy of ${\bf B}^h$ may be written
\begin{align}
W^h &= \int_{S_0}\int_0^h\frac{({\bf B}^h)^2}{8\pi}\,dz'dxdy,\\
&=\frac{1}{8\pi}\int_{S_0}\int_0^T\left[\frac{T}{h}(B^T_x)^2(x,y,z) + \frac{T}{h}(B^T_y)^2(x,y,z)\right. \nonumber\\
&\hspace{3.5cm} \left. +\quad \frac{h}{T}(B^T_z)^2(x,y,z)\right]\,dzdxdy.
\label{eqn:Wh}
\end{align}

\begin{figure}[h]
\begin{center}
\includegraphics[width=\columnwidth]{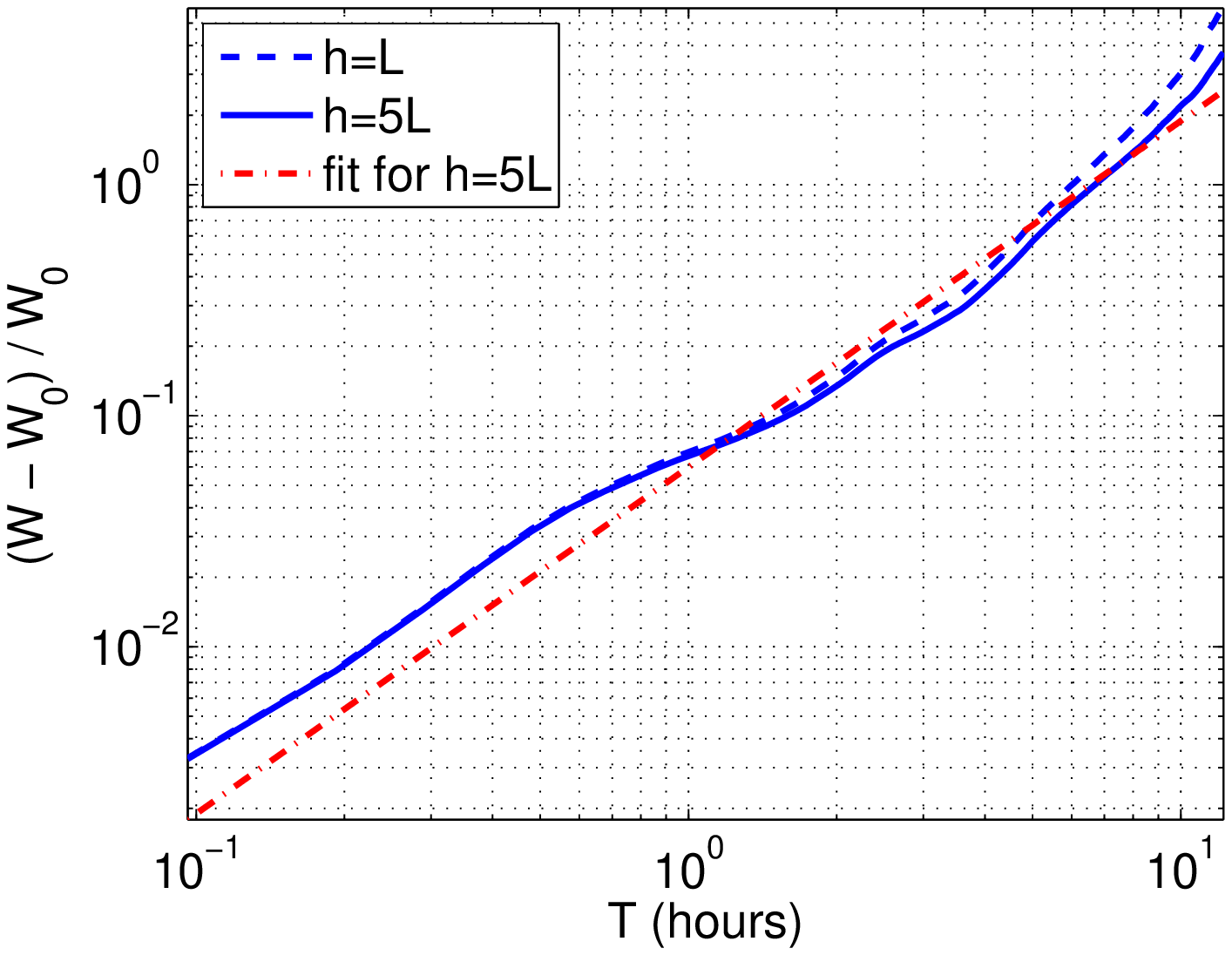}
\includegraphics[width=\columnwidth]{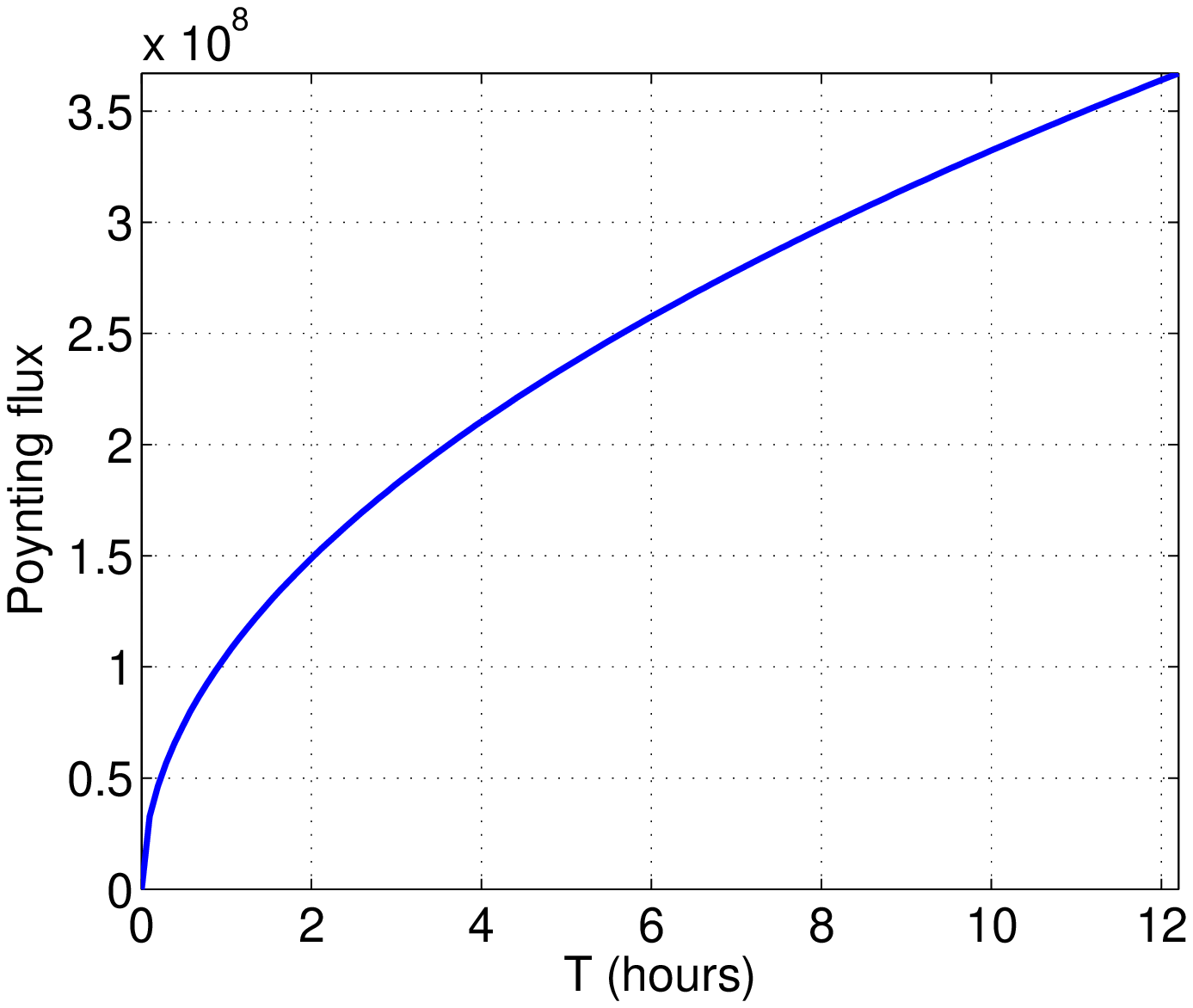}
\caption{Energy estimate from the field reconstruction in Section \ref{sec:recon}. The upper panel shows the relative free energy as a function of $T$ (for two assumed domain sizes $h=L$ and $h=5L$). The lower panel shows the Poynting flux (per unit area) in a domain $h=L$ assuming the fit $at^b$ shown by the dot-dashed line in the left panel.} \label{fig:suspension}
\end{center}
\end{figure}
Since $\lambda(x,y,0)=B_0$, it follows from equation \eqref{eqn:Wh} that the relative free energy $(W^h - W_0)/W_0$ is independent of $B_0$. Evaluating this for our observed ${\bf v}$, we find that the relative free energy is effectively independent of $h$ providing that $h$ is large enough (Figure \ref{fig:suspension}). 

Unlike the lower bounds, the time-dependence of the relative free energy is now super-linear: approximating by a numerical fit with constant exponent (for $h=5L$) gives $W^h/W_0 \sim t^{1.5}$. Notice that this leads to a Poynting flux that increases with time like $dW_h/dt \sim W_0 t^{0.5}$. This is shown in Figure \ref{fig:suspension} for $h=L$ (note that $dW/dt$ scales linearly with $h$ following $W_0$). At $t=12\,{\rm hr}$, this puts an upper bound of about $3.5\times 10^8\,{\rm erg}\,{\rm cm}^{-2}\,{\rm s}^{-1}$ on the Poynting flux.

\subsection{Numerical convection simulations} \label{sec:sim}

As described above, our velocity observations have a rather lower mean speed than expected for granular flows. By comparing with a simple two-dimensional model of convection, \citet{Yeates2012} found that the field line mapping changed significantly if the speed of the photospheric flows was increased. To study the possible change to our Poynting flux estimate that would arise from better-resolved flows, we have also calculated the energy bounds using photospheric velocities taken from a numerical convection simulation \citep{Bushby2012a}.

The horizontal velocities $v_x(x,y,t)$, $v_y(x,y,t)$ have been extracted from the upper boundary of a three-dimensional simulation of hydrodynamic convection in a Cartesian slab $0\leq x\leq 10d$, $0\leq y\leq 10d$, $0\leq z\leq d$, heated from below. Details of the simulations, which are fully compressible, may be found in \citet{Bushby2012a}. Here, the results are dimensionalized by taking $d=1.5\,{\rm Mm}$ and assuming the time scale $t_0=5\,{\rm min}$ (the convective turnover time in the simulation is approximately $3t_0$). In these units, the mean velocity is $\langle(v_x^2 + v_y^2)^{1/2}\rangle\approx 1.7\,{\rm km}\,{\rm s}^{-1}$, comparable to observed granular velocities \citep{rieutord2010}. To prevent particles leaving $V$, we artificially taper $v_x$ and $v_y$ to zero on the side boundaries.

\begin{figure}[h]
\begin{center}
\includegraphics[width=\columnwidth]{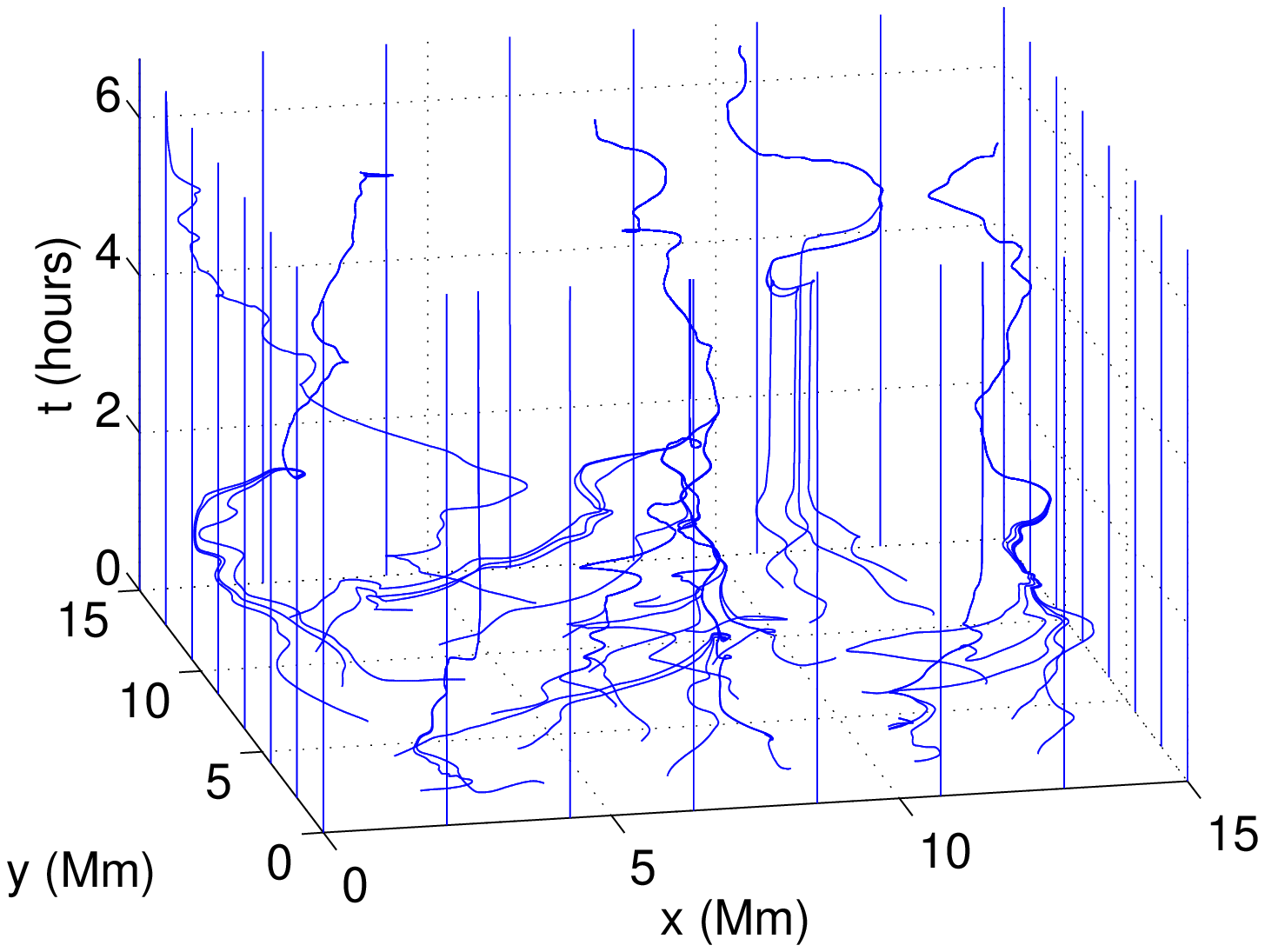}
\includegraphics[width=\columnwidth]{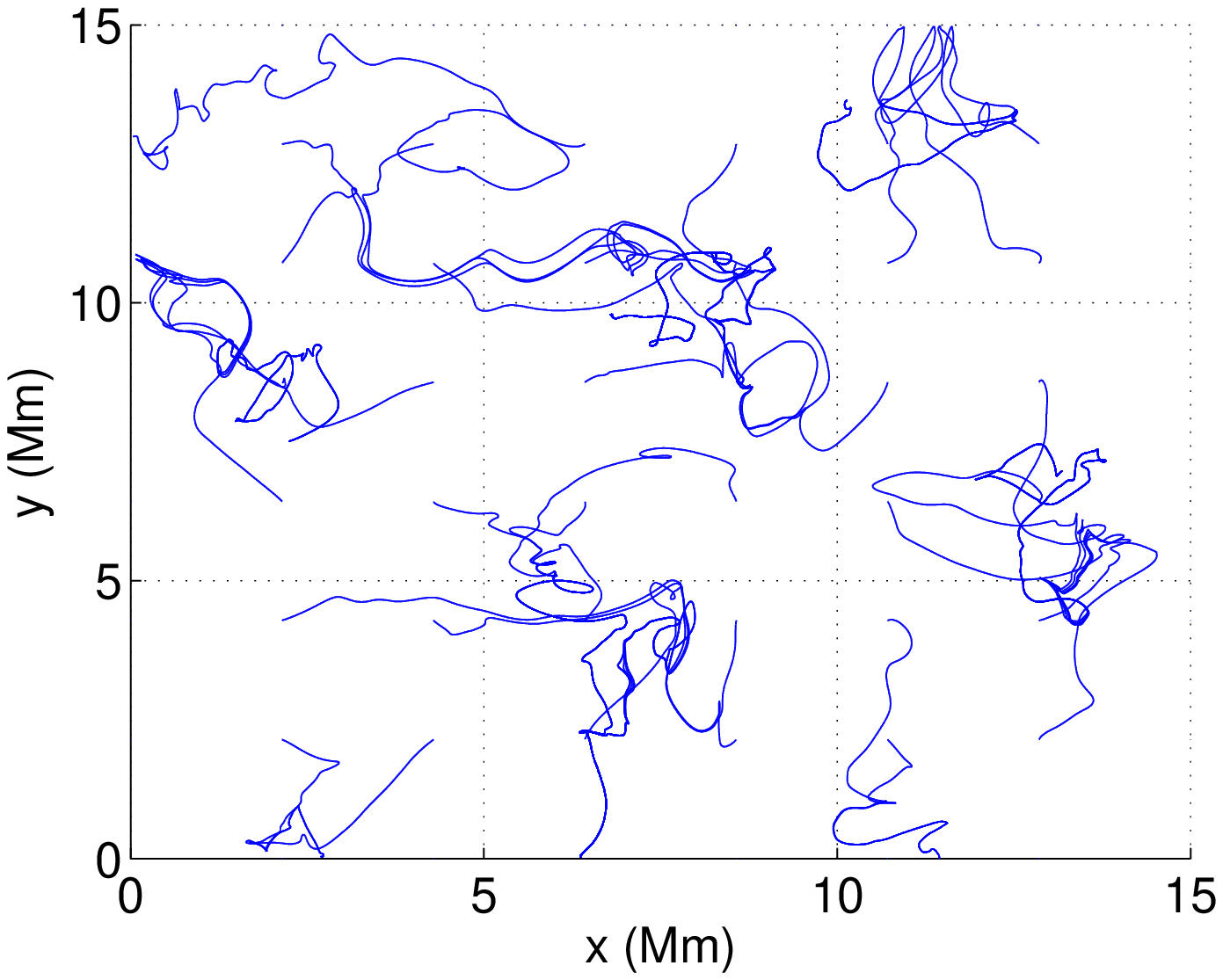}
\caption{Trajectories of the velocity field from the numerical convection simulation, seen from two different angles.} \label{fig:bushby_lines}
\end{center}
\end{figure}

Figure \ref{fig:bushby_lines} shows a typical set of trajectories. Although the domain is similar in size, the trajectories are rather different to those of the observed data (Figure \ref{fig:welsch}). In particular, the faster speed and compressible nature of the flow lead to rapid clumping of trajectories at a small number of locations. But, as with the observed velocities, we find that $W/W_0$ increases approximately linearly in time.

\begin{figure}[h]
\begin{center}
\includegraphics[width=\columnwidth]{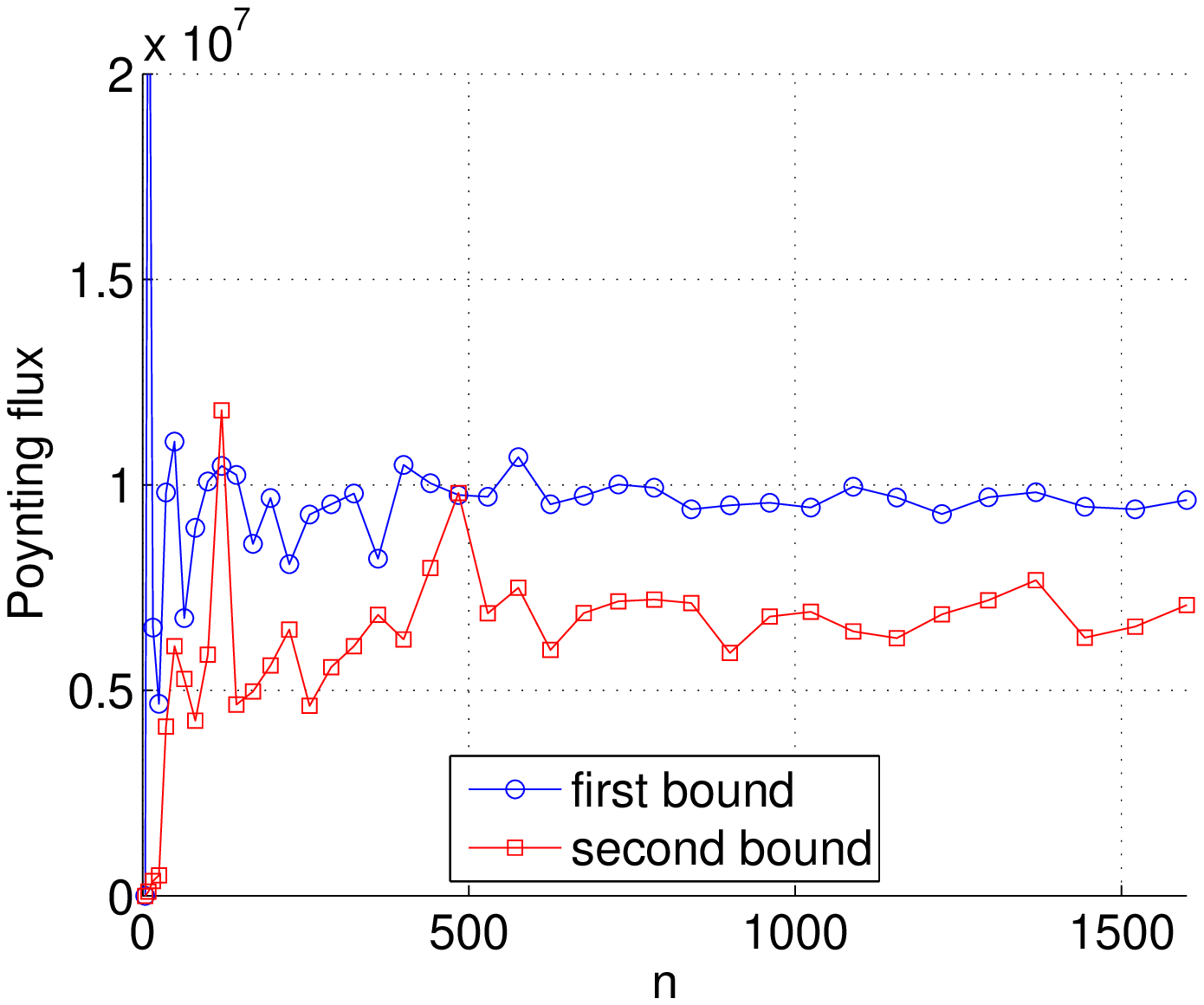}
\includegraphics[width=\columnwidth]{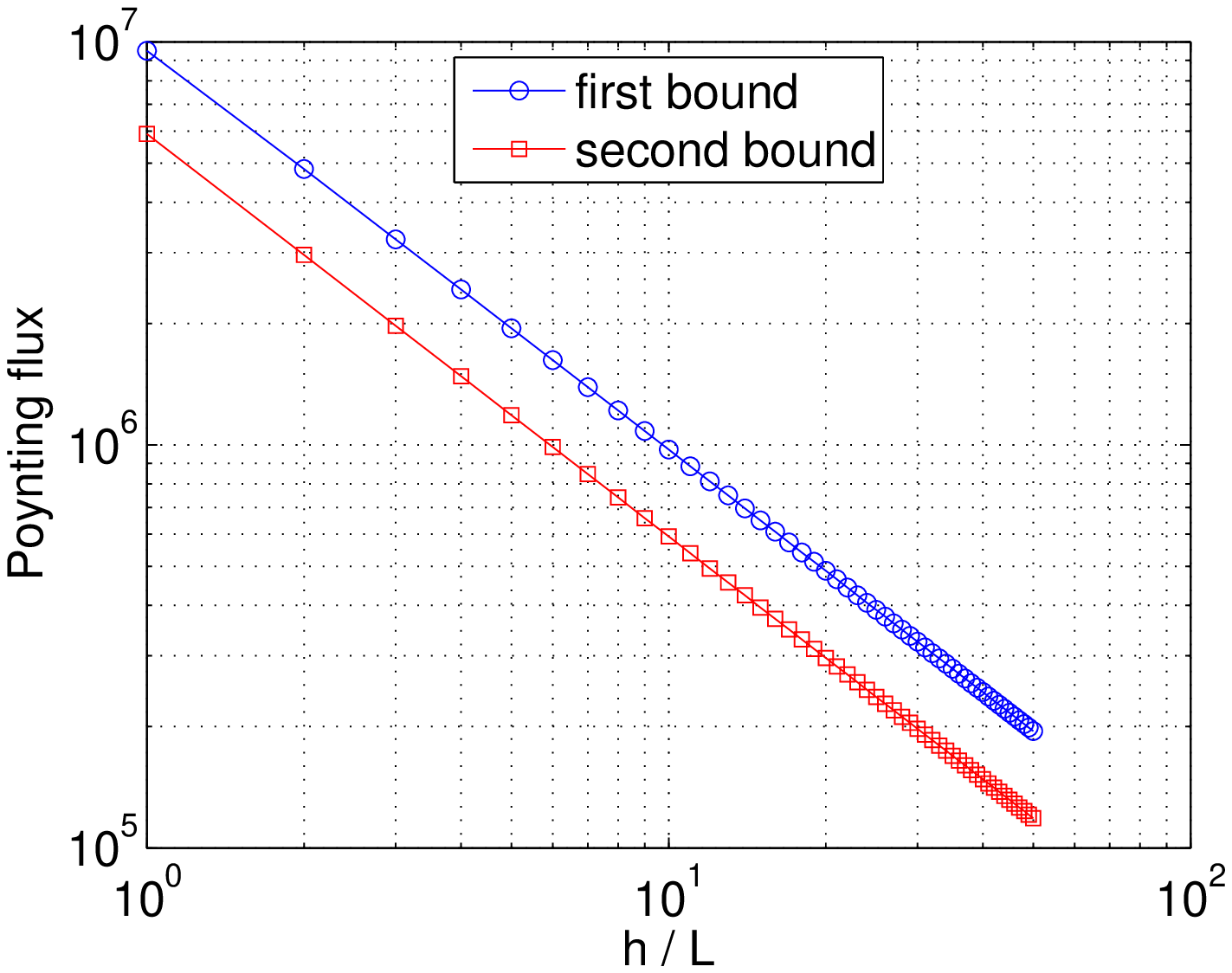}
\caption{Computed energy bounds for the velocity data taken from the numerical convection simulation. The upper panel shows the convergence of the Poynting flux (per unit area) as the number of trajectories used for the calculation is increased (with $h=L$, $B_0=350\,{\rm G}$). The lower panel shows that this estimate scales like $h^{-1}$, as for the observed velocities. Units are ${\rm erg}\,{\rm cm}^{-2}\,{\rm s}^{-1}$.} \label{fig:bushby2}
\end{center}
\end{figure}

\begin{figure}[h]
\begin{center}
\includegraphics[width=\columnwidth]{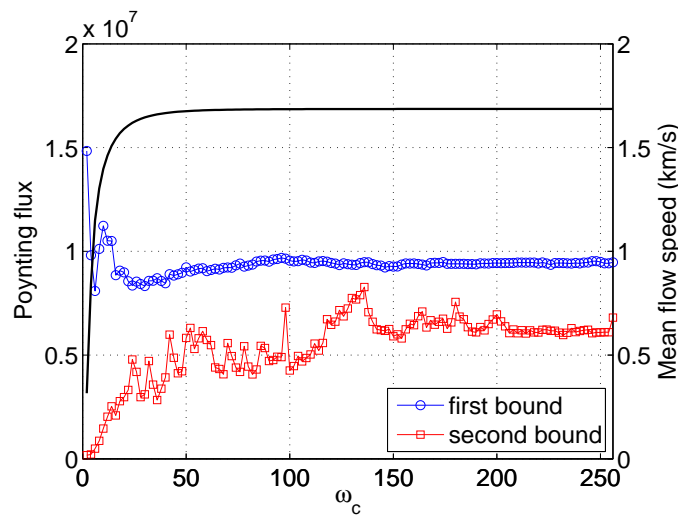}
\caption{Effect of applying a low-pass spatial filter to the velocities from the numerical convection simulation. The horizontal axis shows the cutoff frequency $\omega_c$ of the filter. The solid line shows the mean flow speed $\langle\sqrt{v_x^2 + v_y^2} \rangle$ (right axis), while the two lines with symbols show the two bounds for the Poynting flux in ${\rm erg}\,{\rm cm}^{-2}\,{\rm s}^{-1}$ (left axis). Here we took $h=L$, $B_0=350\,{\rm G}$.} \label{fig:bushby_smooth}
\end{center}
\end{figure}

Estimates of the Poynting flux (for $h=L$ and $B_0=350\,{\rm G}$) are shown in Figure \ref{fig:bushby2}. As before, both bounds scale like $h^{-1}$ as the domain height is increased. For the first bound, we find approximately $1\times 10^7{\rm erg}\,{\rm cm}^{-2}\,{\rm s}^{-1}$. This is only slightly larger than the value calculated for the observed velocity. On the other hand, the second bound is now much closer to the first, at approximately $0.7\times 10^7{\rm erg}\,{\rm cm}^{-2}\,{\rm s}^{-1}$. To explore this difference, we have repeated the calculation for the simulation with varying degrees of smoothing applied to the velocity field before calculating the trajectories. Figure \ref{fig:bushby_smooth} shows the effect on the bounds of applying a low-pass filter of the form
\begin{equation}
G(\omega)=\frac{1}{1 + (\omega/\omega_c)^4}
\end{equation}
in frequency space, where $\omega$ is the spatial frequency and $\omega_c$ is the cut-off. Notice that the first bound is rather insensitive to the filtering out of high-frequency information in the velocity field. On the other hand, the second bound starts to decay for a cut-off frequency as high as 128 (the spatial resolution is $256^2$). Thus it appears that the second bound requires smaller-scale fluctuations in the velocity to measure relative twisting between trajectories. By contrast, the first bound yields similar results even for a rather coarsely averaged velocity field. This finding is consistent with the behaviour of the two bounds for the observations, where the mean flow speed is lower and the velocity is smoother.

\section{Conclusions} \label{sec:conclusions}

In this Paper we have computed two alternative lower bounds for the magnetic energy injected into the solar corona by photospheric footpoint motions, assuming a perfectly ideal coronal evolution. The first bound is based on the displacement between the two footpoints of each field line, and the second bound is based on the relative pairwise twisting of field lines. The advantage of these bounds is that they do not require observations of horizontal magnetic field components in the photosphere, only the initial vertical magnetic field and a sequence of horizontal velocities. We have computed the bounds for an observed sequence of photospheric velocities derived from correlation tracking in Hinode/NFI line-of-sight magnetograms.

For the observed data, we find that the first lower bound is approximately 100 times larger than the second. If the height $h$ of the domain is assumed equal to its horizontal extent $L\approx 12\,{\rm Mm}$, then the first bound gives a Poynting flux of $10^7\,{\rm erg}\,{\rm cm}^{-2}\,{\rm s}^{-1}$, which is roughly equal to the observed coronal heating rate. However, it is probably more realistic to take a longer domain, say $h=8L$, in which case the estimated Poynting flux is approximately $10^6\,{\rm erg}\,{\rm cm}^{-2}\,{\rm s}^{-1}$. A possible reason for this shortfall, compared with Parker's simple estimate of $10^7\,{\rm erg}\,{\rm cm}^{-2}\,{\rm s}^{-1}$ \citep{Parker1983ar}, is the slower flow speeds in our observations. However, when we compute the bounds for a numerical convection simulation with faster flows (Section \ref{sec:sim}), the first bound is only slightly increased. On the other hand, the simulations do suggest that smoothing of the observed velocity field -- mainly due to resolution limits of the observations -- could be responsible for the discrepancy between the second bound and the first. This indicates that braiding of flux ropes may be a more important source of energy than the observations suggest.  On the other hand, both cases show a roughly linear increase in energy with time. \citet{Berger1991q} suggests that such a linear increase is expected if energy is injected mainly by translations of individual flux tubes, and that if energy were mainly injected by entanglement of multiple tubes, a quadratic increase would be expected (as in our simple example in Section \ref{sec:test}).

We have also put a (non-rigorous) upper bound on the magnetic energy by constructing a magnetic field whose field lines are effectively the footpoint trajectories (Section \ref{sec:recon}). At the end of the 12-hour dataset, this gives a Poynting flux of $3.5\times 10^8\,{\rm erg}\,{\rm cm}^{-2}\,{\rm s}^{-1}$, essentially independent of $h$. At one particular time during our observations, we were able to compare these estimates against a Poynting flux estimate using horizontal magnetic field components from a Hinode/SP vector magnetogram. Reassuringly, the estimate of $1.67\times 10^7\,{\rm erg}\,{\rm cm}^{-2}\,{\rm s}^{-1}$ falls between our lower and upper bounds.

Our assumption of a Cartesian domain, and the lack of information on motions at the opposite end of the field lines, mean that our quantitative estimates can only be taken as indicative. And even if the coronal domain were truly Cartesian, there is no guarantee that our lower energy bounds are tight, in the sense of being attainable by relaxation of the magnetic field without allowing reconnection. However, any more detailed estimate would require a model (or observations) of the three-dimensional evolution of the magnetic field in the coronal volume. Such a model would also be required in order to determine when the energy input saturates due to non-ideal dynamics in the corona. It is encouraging that MHD models, which do include this saturation effect - albeit at the expense of an artificially high non-ideal dissipation in the corona - find heating rates comparable to our estimated Poynting flux.

Whilst chosen to match the observed average vertical magnetic field, our assumption of an initially uniform field at the start of the footpoint motions does not account for the concentration of magnetic footpoints in intergranular lanes. Since vorticity is known to peak in the intergranular lanes \citep{Wang1995}, this might affect the injected energy. However, the estimated Poynting flux in the simulations -- where the concentration of footpoints is clearly evident -- is comparable to the observations even after concentration has taken place. Such concentration of field line footpoints must be accompanied by an expansion of flux tubes as they pass through the chromosphere to fill the corona. This expansion will not in itself change the connectivity of field lines, but it is likely to impact on the pattern of energy release in the corona \citep{vanBallegooijen1998d}. Since our study concerns only the build-up of energy, this and other aspects of the energy release remain for further investigation.

\begin{acknowledgements}
We are indebted to Jean-Jacques Aly for sharing pre-publication details of his energy bounds, and to the Royal Astronomical Society for a grant enabling the summer project of FB. ARY was supported by STFC consortium grant ST/K001043/1 to the universities of Dundee and Durham, and thanks E. DeLuca whose question motivated this investigation. Hinode is a Japanese mission developed and launched by ISAS/JAXA, with NAOJ as domestic partner and NASA and STFC (UK) as international partners. It is operated by these agencies in cooperation with ESA and NSC (Norway).
\end{acknowledgements}

\bibliographystyle{aa}
\bibliography{yeates}

\end{document}